# Compression behaviour and crashworthiness analysis of aluminum foam filled corrugated and tapered tubes with graded thickness

Santhosh Reddy[a*], Vignesh S[b], C. Lakshmana Rao[b]

[a,b]Department of Applied Mechanics, Indian Institute of Technology Madras, Chennai-600036, India

[*]*Corresponding author: +91-442 2574059 (Ph), +91-442 22574052 (Fax),  santhosh49reddy@gmail.com*

## Highlights

- Compression behavior of corrugated tubes and tapered tubes with graded thickness were investigated under impact loading.
- Initial peak force and fluctuation in the force-displacement curves of corrugated tube is considerably less when compared to conventional straight tube.
- Tapered tubes with graded thickness is superior to uniform thickness in terms of crashworthy parameters.
- In foam filled tubes, the mode of deformation changes from mixed or diamond to concertina, which is useful in crashworthy applications.
- Foam filled tubes have higher crushing force efficiency than empty tubes in types of tube configurations.

## Abstract

Thin-walled straight circular tubes (SCT) are frequently used as energy absorbing devices in the crashworthy applications. This paper introduces a various tubal configuration, namely aluminium foam filled corrugation tube and tapered tube with graded thickness, to control the collapse mode, and minimize the peak crushing force and fluctuations in force-displacement curves. Dynamic crushing simulations were carried out using commercially available finite element package ABAQUS explicit 6.13 at impact velocity of 60 km/h (corresponding to 16.7 m/s). A comparative study on the dynamic crushing behaviour of aluminum foam-filled tapered with graded thickness and corrugated tubes were performed. The results showed that deformation mode of corrugated tube is more controllable and predictable in the case of empty tubes. In foam filled tubes the mode of deformation changing from diamond or mixed mode to concertina mode which is useful for crashworthy applications. The crushing force efficiency of foam filled tubes increases when compared with the empty tubes because of a higher mean force which can be achieved by less fluctuations in force- displacement curves. The effects of corrugation wavelength and amplitude of corrugation tubes on the collapse mode, peak crushing force and energy absorption were studied. Compared to the conventional straight circular tube, initial peak force and fluctuation in the force-displacement curves of corrugated tube is considerably less. Desired crashworthy characteristics can be obtained by changing corrugation wavelength and amplitude of corrugated tubes.

## Nomenclature

| L | Length of the tube | $F_{max}$ | Initial peak force |
|---|---|---|---|
| D | Mean diameter | $F_{mean}$ | Mean force |
| $R_i$ | Inner radius | $E_{total}$ | Total energy absorption |
| t | Thickness of tube wall | $L_c$ | Crush length |
| V | Velocity of top plate | m | Mass of tube |
| $\nabla$ | Corrugation length | $\eta_c$ | Crushing force efficiency |
| δ | Corrugation amplitude | | |

1. Introduction

There is growing demand for aluminum alloys in automotive, aerospace applications because of the need for light weight structural components and also increasing interest of environmental and safety regulations [1-3]. These alloys are widely used in crashworthiness applications because of its higher energy absorbing capacity during impact loading [4]. Crashworthy structure is required to dissipate impact energy with controllable deformation without instant failure [5]. Mostly thin walled tubes are used as crashworthy structure because of higher energy absorbing capacity and ease of manufacturing. Higher energy absorbing capacity of thin walled tubes depends mainly on the progressive and controllable deformation modes. Many researchers have worked with the aim of improving stabilization of the collapse process and reducing its peak load in order to achieve crashworthy design of moving structures [6].

Within the few past decades, in order to improve crashworthiness of thin walled tubes, several design modifications in the tube have been proposed and developed. Among those, filling cellular materials such as metallic foam [7-10], honeycombs [11], and synthetic foam [12-14] inside the tubes and generating grooved surface on the tubes [15] were found to be efficient way to increase structural safety and energy absorption capacity. In order to improve the collapse mode stabilization and reduction of peak load several investigations have been performed with aim of designing crashworthy structures. Energy absorption of the tubes can be increased by allowing the tubes to deform in concertina mode rather than diamond or mixed mode [16-18]. Alavi et al.[19] investigated the energy absorption capacity of different geometric structure such as circular, square, rectangular, hexagonal, triangular, pyramidal and conical tubes. The authors have concluded that the circular tube is having the highest energy absorption and mean force, while conical tubes are having less difference in initial peak and mean forces. El-Hage et al. [20] studied the effect of triggering, chamfering, hole pattern on aluminum tubes using numerical simulation. They have concluded that triggering can control the folding initiation force, but the mean force was relatively small when compared to conventional straight circular tubes. The literature review shows that the circular tube is having more energy absorption but it generates excessive initial peak force. In many situations, during the collapse process the crashworthy structure has to dissipate the maximum kinetic energy in a stable manner. High initial peak force should be avoided to reduce the large deceleration, which increases the probability of damage and passenger injury. There have been many studies to reduce the initial peak force and strengthen the stability of crushing process. Singace et al.[21] conducted experimental study to find the energy absorption capacity of corrugated tubes. Their experimental results showed that corrugated tubes improve the uniformity of the load-displacement behaviour under axial crushing. Chen et al. [22] investigated the behaviour of cylindrical tubes with corrugated surface subjected axial load using numerical simulation. They have reported that the mode of crushing deformation can be varied with the corrugated shape. Daneshi et al.[23] studied the characteristics of grooved thin walled tubes subjected to quasi static load. This study showed that fluctuations in crushing force can be reduced by grooved surfaces and energy absorption can also be controlled by varying the groove length. Reddy et al. [24] investigated the crushing response of taper tubes of rectangular section under static and dynamic oblique loading. This study showed that taper tubes can withstand oblique impact loads as effectively as axial load. Ahmad et al. [25] examined the energy absorption response of foam filled conical tubes under oblique impact loading. They have reported that foam filled conical tubes increases the energy absorption capacity and resistance to bending. Until now all the studies were carried out with crushing of tapered tubes with constant wall thickness. Recently very few investigations were concentrated on crushing of tubes with variable thickness along longitudinal direction. Chirwa [26] investigated theoretical analysis of inversion buckling of tapered tube with graded thickness along axial direction. He reported that there is significant increase in energy absorption efficiency when compared to the tapered tube uniform thickness.

Collapse mode and crushing behaviour of foam filled tubes and grooved tubes were widely studied and reported, no investigation has been made on closed cell aluminium foam filled corrugated and tapered tubes under impact loading. Motivated by this fact, dynamic crushing behaviour of aluminium foam filled corrugated tubes were investigated in order meet demand of structural safety and increase in energy absorption capacity. Numerical simulations were carried out using ABAQUS explicit 6.13 to obtain detailed information about the crashworthiness parameters of the tubes. In the first section part of the study, crushing simulations of empty and foam filled corrugated 7075-T6 aluminum alloy tubes were conducted at varying corrugation length. In order to have a reference to compare the improvement in the energy absorption capacity simulations were carried out on simple tubes without any corrugations. As follows, crushing behaviour and crashworthiness of empty and foam filled tapered tubes having uniform and graded thickness were investigated and compared. In last part of the study effects of corrugation tube amplitude and corrugation frequency on the crashworthiness of aluminum foam filled corrugated tubes were presented. Crashworthy parameters of all different types of tube were compared and summarized.

## 2. Material modeling and validation

*2.1 Material Model for AA7075 tube*

The material of aluminum alloy 7075 tube was modeled using Johnson-Cook constitutive model [27]. According Johnson-Cook constitutive model, the Von-Mises flow stress ($\sigma$) is obtained as

$$\sigma = \left[A + B\varepsilon_p^n\right]\left[1 + Cln\dot{\varepsilon}^*\right]\left[1 - (T^*)^m\right] \qquad (1)$$

      (a)         (b)        (c)

(a)= Strain hardening effect, (b)= Strain-rate effect, (c)= Thermal softening effect

Where $\dot{\varepsilon}^* = \dfrac{\overline{\dot{\varepsilon}}_p}{\dot{\varepsilon}_D}$ and $T^* = \dfrac{T - T_r}{T_m - T_r}$

$\varepsilon_p$=equivalent plastic strain, $\overline{\dot{\varepsilon}}_p$= Equivalent plastic strain rate(s$^{-1}$), $\dot{\varepsilon}_D$=reference equivalent plastic strain rate(s$^{-1}$), $\dot{\varepsilon}^*$=effective plastic strain rate, $T^*$=homologous temperature, T=absolute temperature (°C), T$_m$=melting temperature of the work material (°C), T$_r$=reference temperature (°C), A=yield strength of the material at reference strain rate and temparature, B=strain hardening coefficient, C=strain rate hardening coefficient, m=thermal softening coefficient, n=strain hardening exponent.

The Johnson-Cook parameters used in current simulation [28] are shown in table 1.

Table 1 Johnson-Cook material model parameters for AA7075-T6

| Constant | A (MPa) | B (MPa) | n | C | m |
| --- | --- | --- | --- | --- | --- |
| AA7075-T6 | 546 | 678 | 0.71 | 0.024 | 1.56 |

*2.2 Material Model for closed cell aluminum foam*

Low density closed cell aluminum foam was modeled by the means of the implementation of an isotropic hardening model contained in the FEA code ABAQUS, which was originally proposed by Deshpande and Fleck [29]. The model assumes the usual decomposition of the total strain rate into its elastic ($\dot{\varepsilon}_{ij}^e$) and plastic part ($\dot{\varepsilon}_{ij}^p$).

$$\dot{\varepsilon}_{ij} = \dot{\varepsilon}_{ij}^e + \dot{\varepsilon}_{ij}^p \tag{2}$$

The elastic part is given by linear constitutive equation

$$\dot{\sigma}_{ij} = C_{ijkl} \cdot \dot{\varepsilon}_{kl}^e$$

Where $C_{ijkl}$ is the elasticity tensor. The yield strength of the foam can be defined by using the potential,

$$\Phi = \hat{\sigma} - \sigma_c \leq 0 \tag{3}$$

Where $\Phi$ is the yield surface, $\sigma_c$ represents the yield strength of the foam in uniaxial compression and $\hat{\sigma}$ is an equivalent stress given by

$$\hat{\sigma} = \frac{[\sigma_e^2 + \alpha^2 \sigma_m^2]}{\left[1 + \left(\frac{\alpha}{3}\right)^2\right]}$$

Where $\sigma_e$ is the von Mises effective stress, $\sigma_m$ is the mean stress. The hydrostatic strength of the foam is given by

$$|\sigma_m| = \left[\frac{1 + (\alpha/3)^2}{\alpha^2}\right]^{\frac{1}{2}} \sigma_c$$

The shape parameter $\alpha$, defines the shape of elliptical yield surface when expressed in $(\sigma_e, \sigma_m)$ space. The incremental plastic strain is given by

$$\dot{\varepsilon}_{ij}^p = \frac{\hat{\sigma}}{H}\left[\frac{\partial \Phi}{\partial \sigma_{ij}}\right] \tag{4}$$

Where H is the hardening modulus. The plastic strain rate $\dot{\varepsilon}_{ij}^p$ is assumed to be normal to the yield surface and hence plastic Poisson's ratio can be specified as a function of $\alpha$.

$$v^p = -\frac{\dot{\varepsilon}_{11}^p}{\dot{\varepsilon}_{33}^p} = \frac{(1/2) - (\alpha/3)^2}{1 + (\alpha/3)^2} \tag{5}$$

To determine the input parameters of the crushable foam model, closed-cell aluminum foam with relative density of 0.16 was processed by Alporas method. In this method Aluminum ingot is melted at 680°C. Then 1.5 wt.% of calcium was added to the melt as followed by stirring for 2 min. Subsequently 1.6 wt.% of powdered $TiH_2$ was added. Decomposition of $TiH_2$ resulted in foam formation. Subsequently the liquid foam was solidified [30]. Ten prismatic foam specimens were cut having a regular square prisms of size 20 mm×20 mm×40 mm were cut. These foam specimens were subjected to uni-axially compression test in a universal testing machine at a cross head speed of 1 mm/min. Young's modulus (E) was measured from the slope of the loading curve, and it was found that the E of aluminum foam of relative density 0.16 to be 0.7 GPa. Olurin et al has reported that young's modulus of Al foam of relative density 0.15 was found to be 0.78 GPa from unloading curve [31], which is similar to the value obtained in this study. The other measured mechanical properties of foam were similar to that reported by Olurin et al [31].

Table 2 Mechanical properties of aluminum foam under quasi static compression.

| Foam Density, $\rho^*$ (g/cm$^3$) | Relative density, ($\rho^*/\rho_s$) | Young's Modulus, E (GPa) | Plateau stress, $\sigma_{pl}$(MPa) | Densification strain, $\varepsilon_D$ |
|---|---|---|---|---|
| 0.4191 | 0.16 | 0.7 | 2.5 | 0.57 |

*2.2.1 Rate dependence on closed cell aluminum foam core*

At higher strain rate, aluminum foam shows an increase in the yield stress. Cowper-Symonds overstress power law [32] was implemented to define strain rate dependence. Based on power law the dynamic flow stress is expressed as

$$\bar{\sigma}^c_{dyn} = \bar{\sigma}^c_{sta}\left[1+\left(\frac{\bar{\varepsilon}^p}{d}\right)^{\frac{1}{n}}\right]$$

(6)

Where $\bar{\sigma}^c_{sta}$ is the static uniaxial compression yield stress, $\bar{\sigma}^c_{dyn}$ is the yield stress at a non-zero strain rate and $\bar{\varepsilon}^p$ is the equivalent plastic strain rate. Strain rate sensitivity n=1.285, d=2319 1/sec calculated from dynamic compression test was used in simulation.

*2.3 Validation of material model*

In order to validate the Crushable foam and Johnson Cook material model a dynamic compression test was conducted using Split Hopkinson Pressure Bar (SHPB) technique. The tests were conducted on rectangular aluminum foam specimen of size 13 mm × 13 mm × 10 mm and AA7075 tube specimen of 30 mm length and 20 mm inner diameter and 1.5 mm thickness (see Fig 1b). The aluminum bars of diameter 19.5mm and maraging steel bars of diameter 50mm were used for foam and tube specimens respectively.

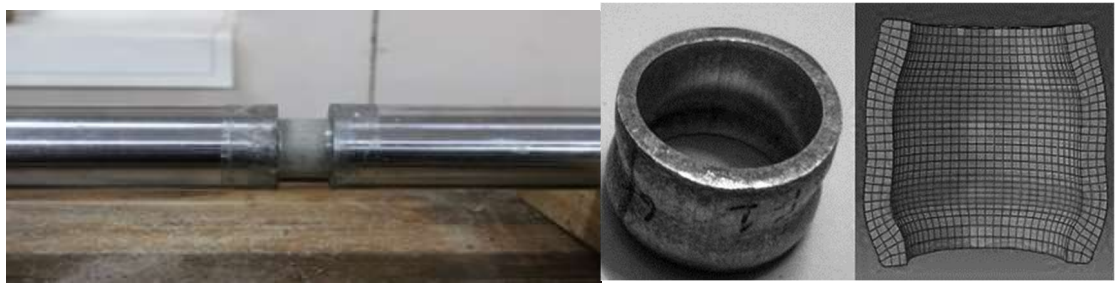

Figure 1. (a) Experimental setup (b) Experiment and simulation comparison of deformed shape of AA7075 tube.

Finite element model was built according to SHPB test procedure [33]. The striker, incident and transmitted bar were modeled using 8-node linear brick elements (C3D8R available in ABAQUS Library) with reduced integration and hourglass control (see Fig 2). All the contacting surfaces between bars were defined as frictionless contact. Both static and dynamic friction coefficient was set to 0.2 on contact surfaces between specimen and bars, as it has little influence on the simulation results [34]. The different incident velocities were set to the striker by means of a predefined velocity field. Initial boundary conditions were applied such that the striker, the incident bar and the transmitted bar can only move in one direction. The nominal stress and nominal strain at strain rate of 1200 s-1 and 2800 s-1 obtained from simulations were validated with the experimental results, as shown in Fig 3 and Fig 4. Good agreement indicates the accuracy and reliability of the FE model.

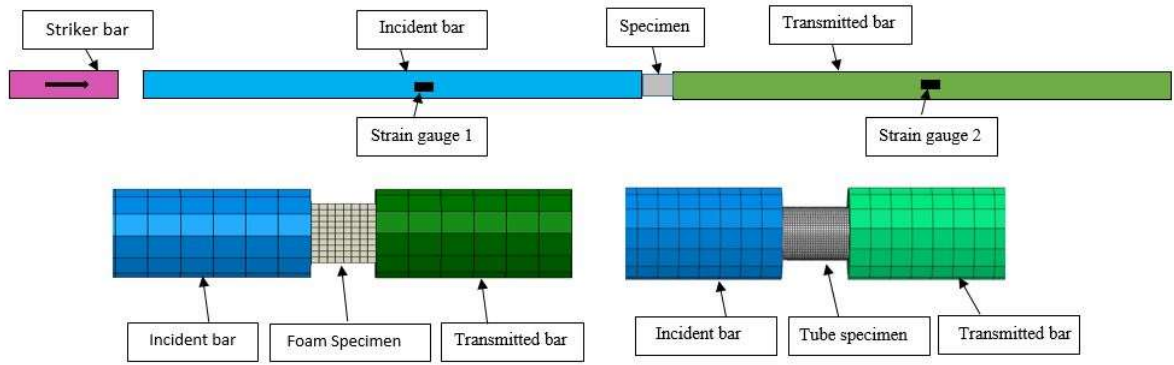

Figure 2. Schematic of the SHPB test arrangement and detail of the model assembly

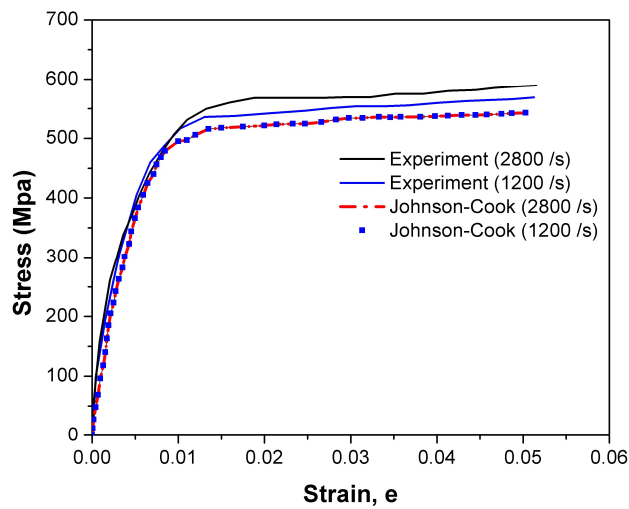

Figure 3. Comparision of J-C model with experimental results

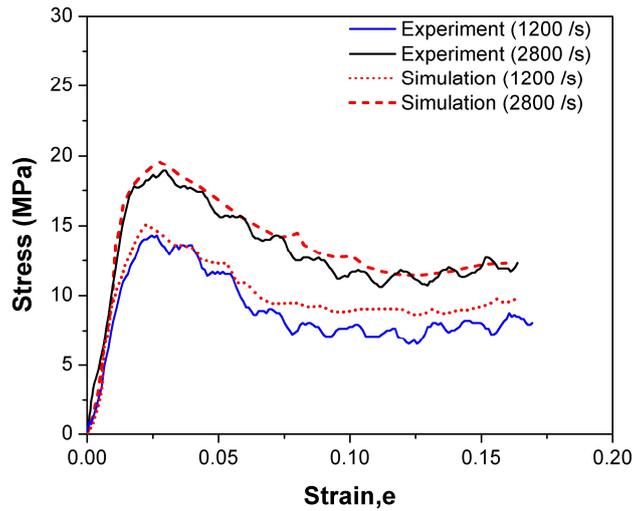

Figure 4. Comparision of dynamic response of foam between experiment and simulation

## 3. Finite element modeling and validation

*3.1. Finite element model*

Numerical simulations were carried out using the commercially available FE package ABAQUS Explicit 6.13 to compare the effect of filling foam inside circular, corrugated and tapered tube in terms of crashworthiness parameters. AA7075 tubes and foam were modeled using the 8-node solid elements (C3D8R) with reduced integration and in combination with hourglass control available in ABAQUS library. Element size of 2 mm was chosen for both tube and foam respectively. The fixed and moving rigid plates were modeled using discrete rigid elements. The finite element model consists of four parts (Tube, foam, top and bottom rigid plates). Two reference nodes were created on both top and bottom rigid plates. The top plate reference node was used to specify values of impact velocity, mass and for measuring compressive deformation of tubes and the bottom plate reference node was used for measuring the reaction force. The empty and foam filled tube models were set between two rigid plates, and the top plate moving toward the bottom plate at a velocity, as shown in Fig.5. Both static and dynamic friction coefficient was set to 0.2, which has small influence on the simulation results [33].The bottom side plate was fixed in all directions and top plate was allowed to move in the y-axis only.

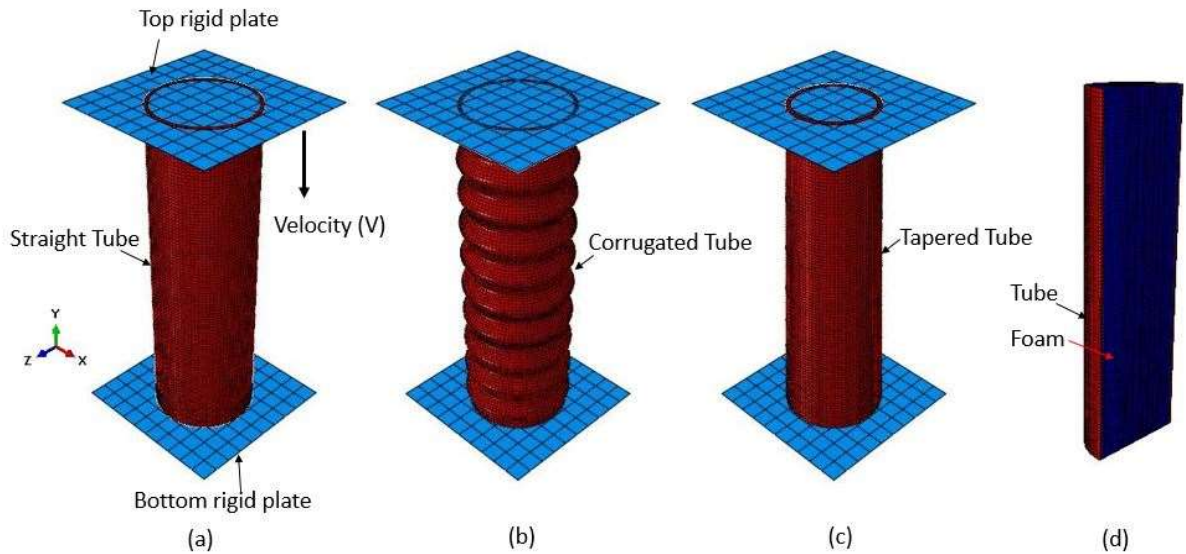

Figure 5. Numerical model of crushing of (a) Straight tube, (b) Corrugated tube, (c) Tapered tube and (d) Sectional view of foam filled tube

The numerical model of the AA7075 straight circular tube is shown in Fig. 5(a). The straight tube has the dimension of 200 mm length, 58.5 mm in inner diameter and 1.5 mm in thickness. This crashworthy structure is used in the front side rail of a passenger car, the length of frontal crash box is often 100-200 mm, so in this paper we choose L=200 mm as the total length of this tube.

In corrugated tubes, Corrugations were created in sinusoidal form of corrugation length (v) and amplitude (δ). In this study, corrugated tubes corrugation amplitude (δ) was first set to 2 mm while corrugation length (v) was varied from 7, 10 and 15 mm. The beginning value of corrugation length 7 mm was equal to folding length of compressed empty straight circular tube.

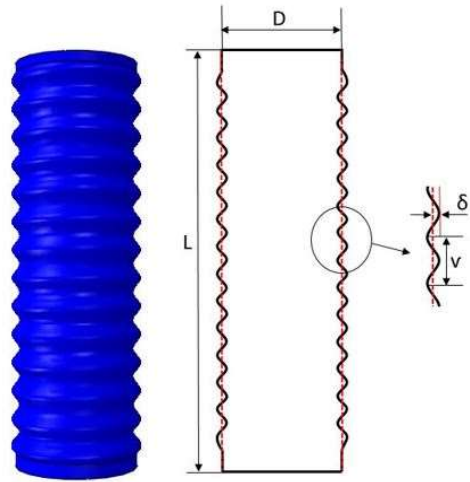

Figure 6. Schematic diagram of corrugation profile

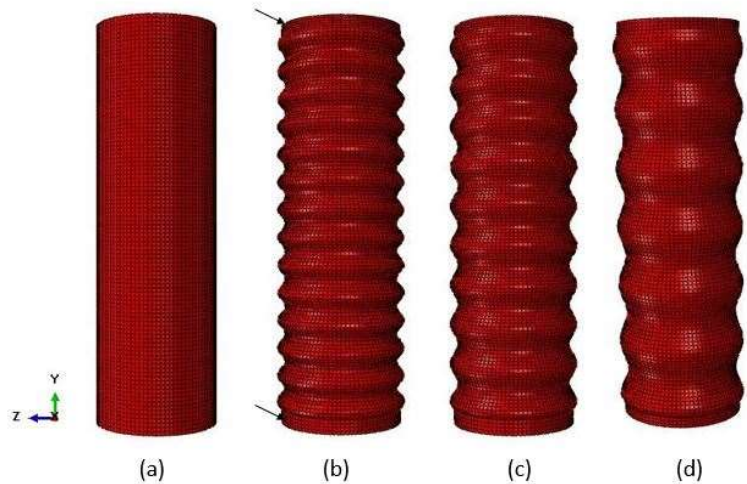

Figure 7. Schematic view of empty tubes; (a) Straight, (b) Corrugated tubes with v of 7 mm, (c) 10mm, (d) 15 mm.

Fig. 7(b),(c),(d) shows the numerical model of empty corrugated tubes. The shown corrugated tubes having amplitude of 2 mm and corrugation length of 7, 10 and 15 mm. Each corrugated tube has straight portions at both the ends as shown in Fig.7(b) with arrows. The length of straight portion at the top end of the corrugated tube was set to be 5 mm while the bottom end varies between 5 and 10 mm according to the corrugation length.

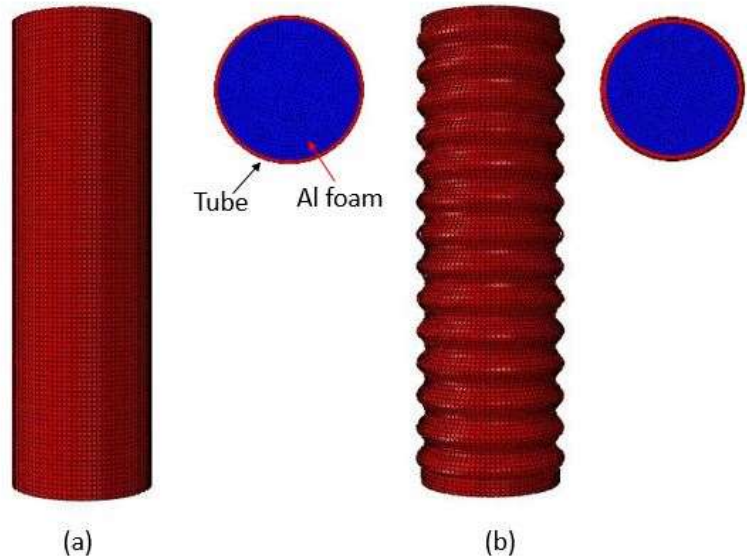

Figure 8. Schematic view of Aluminum foam filled tubes; (a) Straight, (b) Corrugated tubes with v of 7 mm.

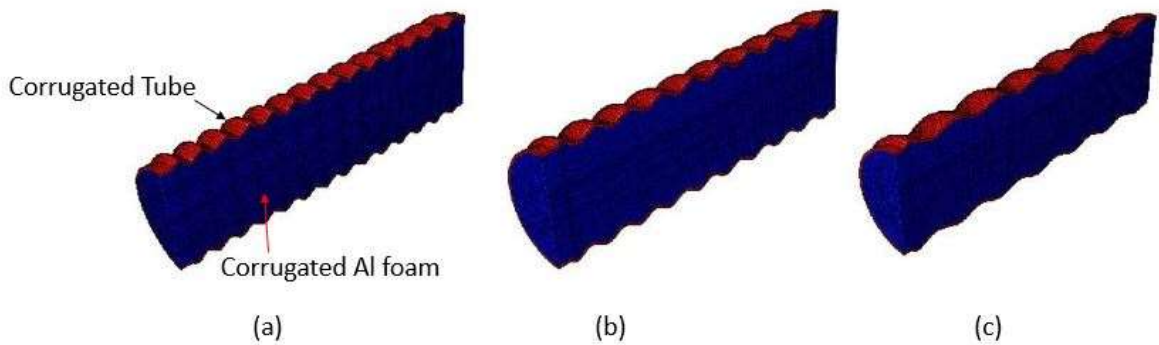

Figure 9. Foam filled tubes with corrugated foam of v of (a) 7 mm, (b) 10mm, (c) 15 mm.

For an energy absorber maintaining load uniformity is another important index besides with the SEA value. Because of variation in diameter of tapered tubes, load uniformity of this tube is lesser than that of straight circular tube. Load fluctuations in tapered tubes between different folds can be eliminated by varying the graded thickness in the two ends of a tapered tube. Wierzbicki T et.al [35] reported that the mean force of two end foldings will keep constant if

$$D_1{}^{0.5}t_1{}^{1.5} = D_2{}^{0.5}t_2{}^{1.5} \tag{7}$$

Finite element model of tapered circular tubes with uniform and graded thickness by maintaining same smaller and larger end diameters in both cases as shown in Fig. 10 (a), (b)

The larger, smaller end inner diameters of tapered tubes are 58.5 mm and 46.5 mm respectively.

Fig.10 (a) shows the sectional view of the DETU of 1.5 mm throughout its length. While the thickness of larger and smaller end of DETG is 1.5 mm and 2 mm respectively. Fig.10 (b) shows the sectional view of DETG.

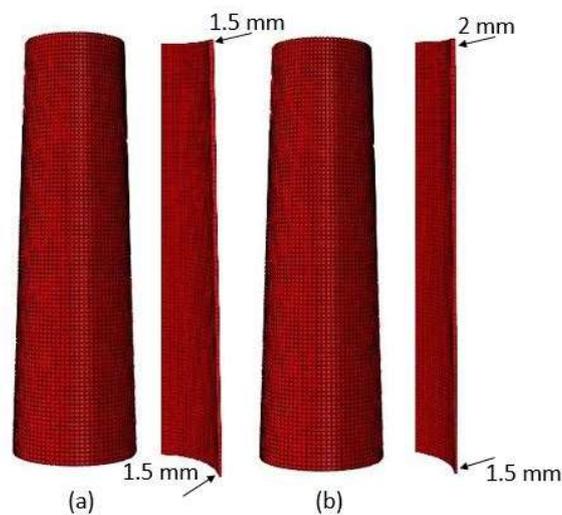

Figure 10. Schematic view of tapered tubes; (a) uniform thickness, (b) graded thickness

In all simulations, the axial impact velocity of 16.7 m/s was applied to the top rigid plate and mass of the rigid wall is set to be 275 kg. The contact interaction between Aluminum foam and tubes were modeled using Automatic surface to surface contact with a friction coefficient of 0.2.

*3.2. FE model validation*

Corrugated will become a straight circular tube when amplitude δ=0 mm, hence the straight tube can be considered as a special case of corrugated tube. Thus the FE model was validated using a straight circular tube with the diameter of 4.5 mm, length of 11.8 mm and thickness of 0.127 mm under axial compression at a constant velocity of 0.6 mm/min. The comparison of force-displacement curve from the numerical simulation with experimental results taken from [36], is presented in Fig. 11. The collapse mode from the FE simulation agrees well with that from the experiment. For the total energy absorption, the numerical and experimental results are 1406 J and 1268.85 J, respectively, which are also in reasonable good agreement.

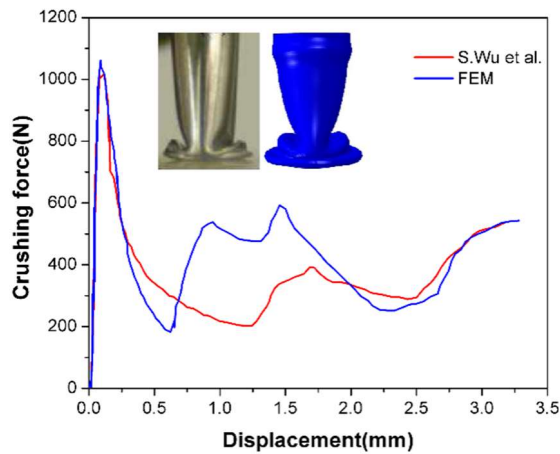

Figure 11. Schematic view of tapered tube

## 4. Crashworthiness Parameters:

In the design of Crashworthy structures, it is essential for the design engineer to know different crashworthy parameters in detail. As a result, different crashworthy parameters are defined to make it easy to compare different crashworthy structures. These parameters are illustrated in the following section.

*4.1 Initial peak force ($F_{max}$)*

Generally, during beginning of crushing of a thin walled tube experienced by a maximum force. The force at this point is called as Initial peak force ($F_{max}$). As a device for energy absorption, the initial peak force should be reduced or restricted to a safe level to reduce the severe injury or damage.

*4.2 Mean crushing force ($F_{mean}$)*

Mean crushing force is the average value of crushing force of the energy absorbing structure through complete deformation. This parameter plays a major role in performance measurement of energy absorbers.

$$F_{mean} = \frac{E_{total}}{L_c} = \frac{\int_0^{L_c} F(x)dx}{L_c} \qquad (8)$$

Where $E_{total}$ is the total energy absorption and $L_c$ is the crush length

*4.3 Total Energy absorption (TEA)*

The amount of energy absorbed throughout the crushing process is called total energy absorption. Energy absorbed at any moment can be calculated by taking area under Force-displacement curve.

$$E_{total} = \int_0^{L_c} F(x)dx \qquad (9)$$

*4.4 Specific Energy Absorption (SEA)*

It is defined as ratio of energy absorbed to mass of tubular structure. This parameter can be used to compare the energy absorption capacity of different material and structures.

$$SEA = \frac{E_{total}}{m} \qquad (10)$$

Where m is mass of tube

*4.5 Crushing Force Efficiency ($\eta_c$)*

It is defined as ratio of mean force ($F_{mean}$) to initial peak force ($F_{max}$) and it can be calculated using the following equation

$$\eta_c = \frac{F_{mean}}{F_{max}} \tag{11}$$

## 5. Results and discussion

The result of the numerical study carried out using ABAQUS Explicit 6.13 to understand mode of deformation and energy absorption behaviour of corrugated and tapered tubes in comparison with straight tubes are discussed in the following sections.

### *5.1 Mode of deformation and Load-displacement curves of empty straight and corrugated tubes*

The corrugated tubes with different corrugation wavelengths and empty straight tubes were compressed at the velocity of 16.7 m/s. Force–displacement curves of empty straight and corrugated tubes of wave length v = 7, 10 and 15mm is shown in Fig 12. Initially the empty straight tube deformed elastically until about a peak load. Thereafter, the tube plastically collapsed followed by fold formation due to local buckling. At this point, there was shift in deformation mode (compression mode to bending mode). The resisting force of the tube decreased and the plastic fold started to collapse which have caused a reduction in load. When the plastic deformation progressed, the folds came in contact with each other. As soon as folds came in contact with each other, the load increases until the next plastic fold was initiated this entire process is repeated and the folds were formed continuously. The mixed modes were observed in the simulation as shown in Fig 13. The highest and lowest peak forces were obtained as 140 kN and 60kN in empty straight tube. The collapse mode shape of empty straight tube subjected to impact compression at the velocity of 16.7 m/s is shown in Fig 13.

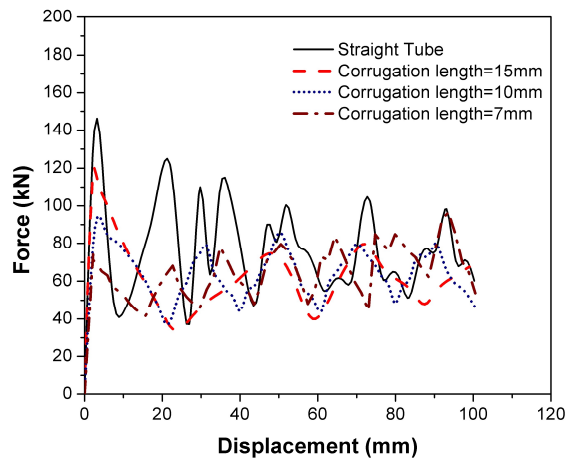

Figure 12. Force –displacement curves of empty straight and corrugated tubes

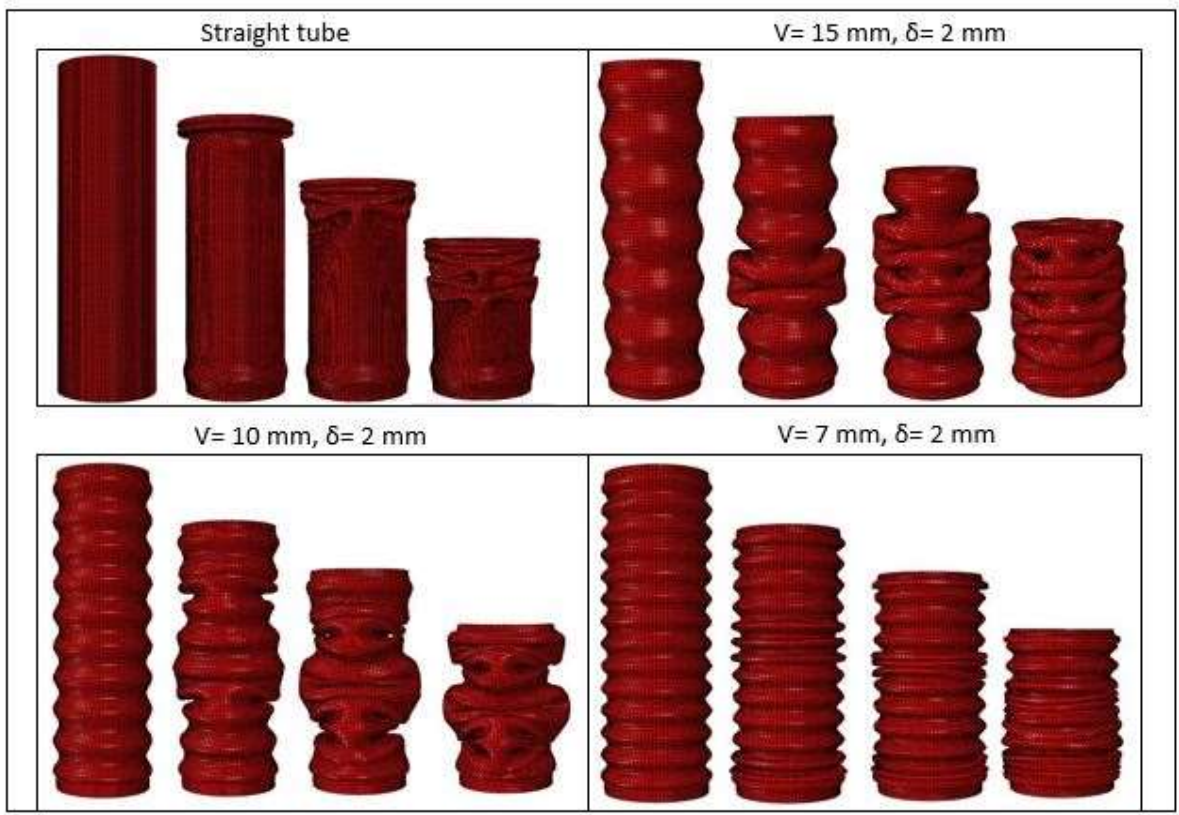

Figure 13. Deformation of empty straight and corrugated tubes at 100 mm displacement

When the corrugation wavelength (v=10, 15 mm) the tube underwent mixed mode of deformation. However, the deformation mode transformed to concertina mode in general, when the corrugation wavelength was less than 7mm. Force-displacement curves of the corrugated tubes were found to be uniform which is favourable characteristic result for designer. However in the corrugated tubes a part of the plastic energy had been consumed during the corrugation fabrication process and hence the mean crush force in the corrugated tubes was smaller than that of the empty straight tube. The initial peak forces of the corrugated tubes of frequencies ν = 7, 10 and 15mm were 71 kN, 87.58 kN and 120 kN, respectively. It was observed that the initial peak forces in force- displacement curves increases with increase in corrugated tube wave length and the mean force also increases with increase in wave length.

*5.2 Mode of deformation and Load-displacement curves of foam filled straight and corrugated tubes*

The collapse mode of foam filled straight tube and corrugated tubes of different corrugation wavelength subjected to impact load at the velocity of 16.7 m/s is shown in Fig 15. As a result in all types of the corrugated and straight circular tubes, the concertina failure mode (axisymmetric) was observed (see fig 15.). The initial peak forces of circular straight and corrugated tubes are shown in table 3. It was found that the peak force in corrugated tubes reduced significantly when compared to the straight circular tube. The initial peak force of straight tube is 157.42 kN which is higher than the corrugated tube peak forces. Overall, around 13-26.9 % of reduction in peak force can be achieved by introducing the corrugations to a conventional straight tube. It was observed that the initial peak forces and the mean forces increases with increase in wavelength of corrugation. Force –displacement curves of foam filled straight and corrugated tubes of frequencies ν = 7, 10 and 15 mm is shown in Fig 14.

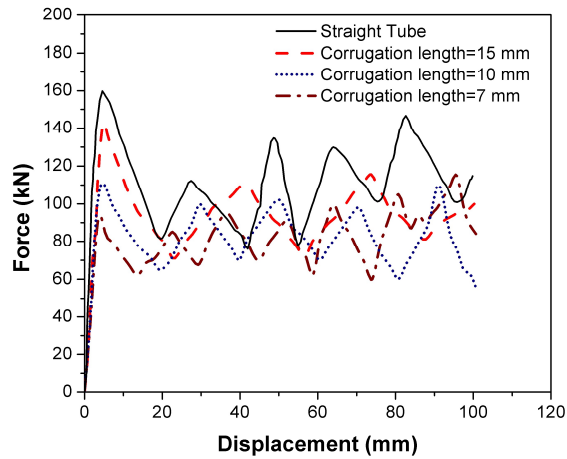

Figure 14. Force –displacement curves of foam filled straight and corrugated tubes

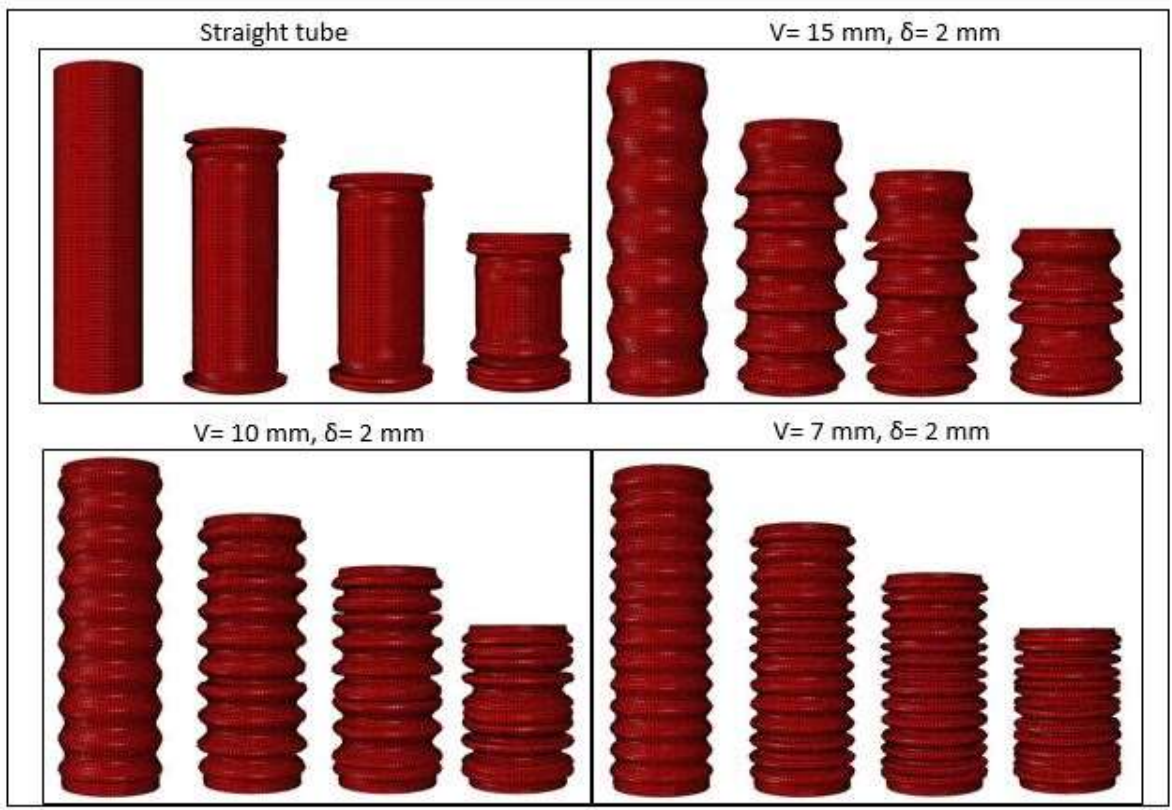

Figure 15. Deformation of foam filled straight and corrugated tubes at 100 mm displacement

*5.3 Mode of deformation and Load-displacement curves of empty straight and tapered tubes*

The Force-displacement response of the empty straight circular and tapered tubes is shown in Fig.16. The initial peak force of graded thickness tapered tube is almost same as that of straight circular tube because of more thickness at the smaller end. In tapered uniform thickness tube the initial peak force was less when compared to straight and graded tubes because the first folding occurred at smaller end. The mean forces for straight, uniform and graded tubes were observed as 105.15 kN, 101.22kN and 123.14kN respectively.

Table 3 Crashworthiness parameters of empty and foam filled tubes

| Type | Specimen no. | mass (gm) | $F_{max}$(kN) | $F_{mean}$(kN) | TEA(kJ) | SEA(kJ/kg) | CFE | Mode of deformation |
|---|---|---|---|---|---|---|---|---|
| Empty Straight tube | DEST | 148.7 | 140 | 75.79 | 8.96 | 60.54 | 0.54 | mixed |
| Foam filled straight tube | DFST | 362.7 | 157.42 | 105.15 | 11.04 | 30.49 | 0.66 | concertina |
| Empty Corrugated tube (v=7mm,δ=2mm) | DEC7 | 172.9 | 71 | 64.19 | 9.38 | 60.9 | 0.61 | concertina |
| Foam filled Corrugated tube (v=7mm,δ=2mm) | DFC7 | 383.9 | 115 | 81.25 | 8.23 | 21.48 | 0.71 | concertina |
| Empty Corrugated tube (v=10mm,δ=2mm) | DEC10 | 164 | 87.58 | 63.01 | 6.42 | 39.15 | 0.72 | mixed |
| Foam filled Corrugated tube (v=10mm,δ=2mm) | DFC10 | 377 | 110 | 80.45 | 8.75 | 23.21 | 0.73 | concertina |
| Empty Corrugated tube (v=15mm,δ=2mm) | DEC15 | 152 | 120 | 63.82 | 6.61 | 43.48 | 0.53 | diamond |
| Foam filled Corrugated tube (v=15mm,δ=2mm) | DFC15 | 367 | 137 | 93.32 | 9.4 | 25.61 | 0.68 | concertina |
| Empty Taper uniform | DETU | 134 | 157.32 | 87.13 | 7.72 | 57.61 | 0.55 | mixed |
| Foam filled Taper uniform | DFTU | 288 | 157.5 | 101.22 | 9.91 | 32.38 | 0.64 | concertina |
| Empty Taper graded | DETG | 154 | 158 | 96.4 | 9.38 | 60.9 | 0.61 | mixed |
| Foam filled Taper graded | DFTG | 323 | 171.6 | 123.14 | 12.72 | 39.38 | 0.72 | concertina |

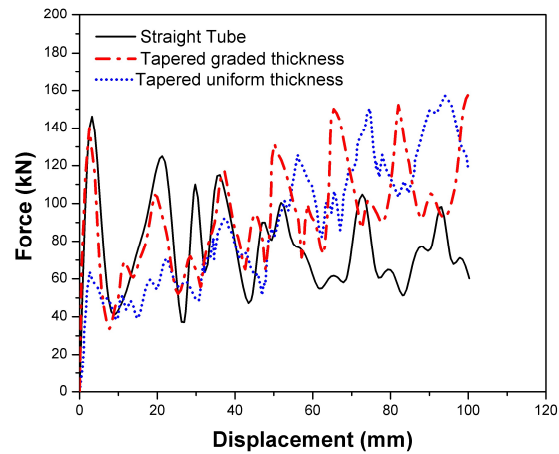

Figure 16. Force –displacement curves of empty straight and tapered tubes

Fig 17. Shows the deformation mode of the tapered tube having uniform thickness under impact loading. The tube initially deformed at the top end because of smaller diameter. First two folds were observed as concertina mode of deformation as shown in Fig 17. After that it transforms to diamond mode in the subsequent progressive deformation. Finally the empty tapered tube having uniform thickness undergone mixed mode of deformation in which diamond mode of deformation occur after few concertina modes which is unfavourable for crashworthy structures. Fig 17 shows the deformation mode of the tapered tube having graded thickness in which the tube initially deformed at the larger end instead of smaller end as observed in uniform thickness tube. In graded thickness taper tube the thickness is varying linearly along the length of tube i.e. smaller end having more thickness when compared to larger end because of this reason deformation occurred at larger end. In tapered tube with graded thickness also first two folds were observed as concertina mode and after that diamond mode in the subsequent steps. The mode of deformation is almost same in both the cases.

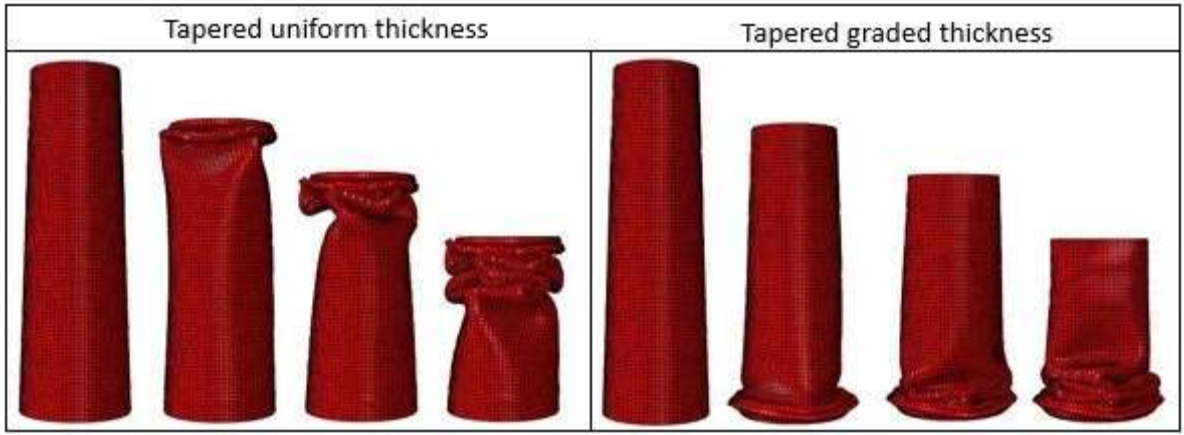

Figure 17. Deformation of empty straight and tapered tubes at 100 mm displacement

## 5.4 Mode of deformation and Load-displacement curves of foam filled straight and tapered tubes

Fig. 18 shows the Force-displacement responses of the foam filled straight circular and tapered tubes. The initial peak load of all foam filled tubes were higher than that of corresponding empty tubes, this is because of the tube foam interaction which increases the strength of the tube. In straight circular tube the force drops and fluctuates about a mean force after the initial peak force. But in tapered tubes the force increases after the initial peak force because of its geometry. The maximum peak forces of straight circular, uniform and graded thickness tubes are 157.42 kN, 157.5 kN and 171.6 kN respectively. The mean crushing forces of straight circular, uniform and graded thickness tubes are 105.15 kN, 101.22 kN and 123.14 respectively.

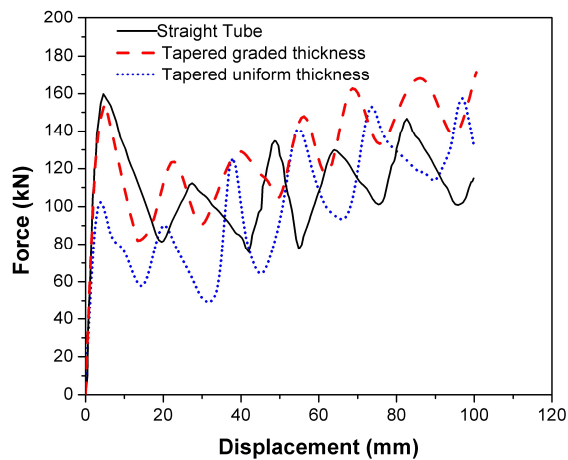

Figure 18. Force –displacement curves of foam filled straight and tapered tubes

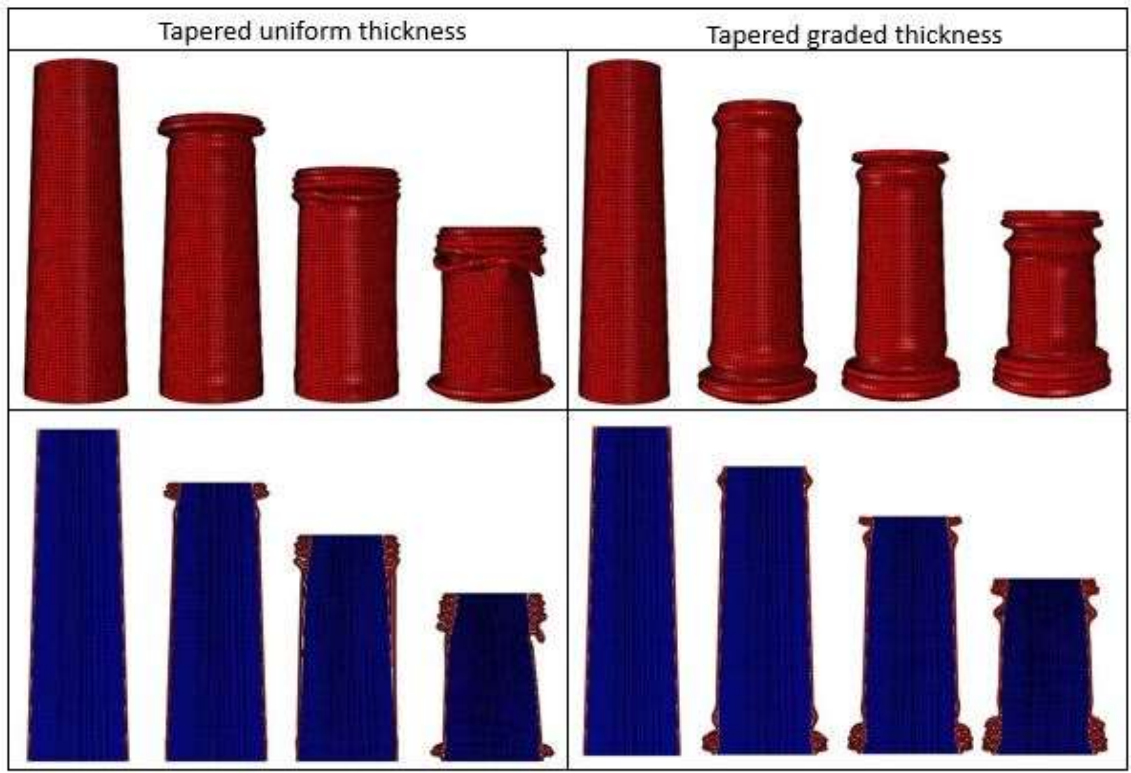

Figure 19. Deformation of foam filled straight and tapered tubes at 100 mm displacement

## 5.5 Specific energy absorption and crash force efficiency

Fig 20(a) shows the SEA and CFE values of empty straight and corrugated tubes of different wavelength. The total energy absorption values of empty straight and corrugated tubes of frequencies v = 7, 10 and 15 mm is shown in table 3. The highest and lowest energy absorption were found to be 8.96 kJ in straight tube and 6.11 kJ in corrugation wavelength of 7mm. The straight tube having highest SEA of 60.54 kJ/kg and lowest was found to be 35.31 kJ/kg in corrugation wavelength of 7mm. The maximum and minimum CFE values were 0.9 in corrugated tube with v=7mm and 0.53 in corrugated tube with v=15mm respectively.

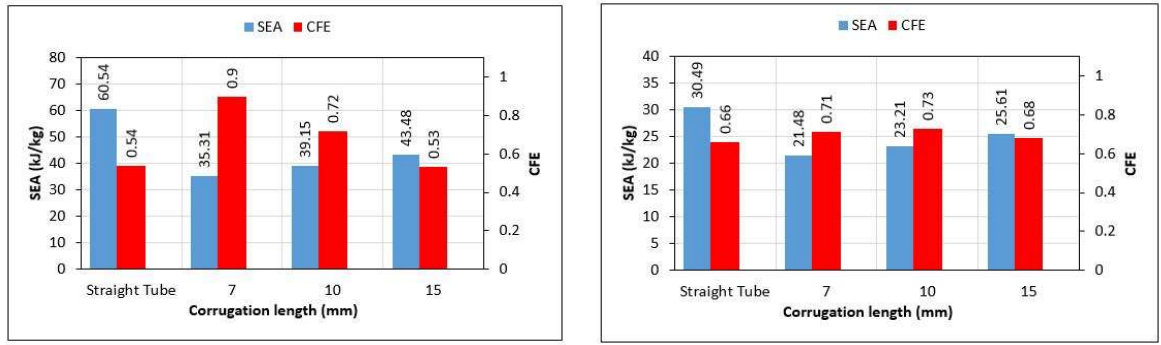

Fig 20. SEA and CFE of (a) empty straight and corrugated tubes, (b) Foam filled straight and corrugated tubes

Fig 20(b) shows the SEA and CFE values of foam filled straight and corrugated tubes. The highest and lowest total energy absorption were found to be 11.04, 8.23 kJ in straight and corrugation tube with v=7mm. Each foam filled tube showed higher energy absorption value than empty tubes. It was found that foam filled straight tube having higher SEA of 30.49 kJ/kg and corrugation tube with v=7mm having lower SEA of 21.48 kJ/kg. The highest CFE value, 0.73 was found in corrugation tube with v=10 mm and lower CFE value, 0.66 found in straight tube. For both empty and foam filled straight and corrugated tubes, CFE value was found to decrease with increasing corrugation length.

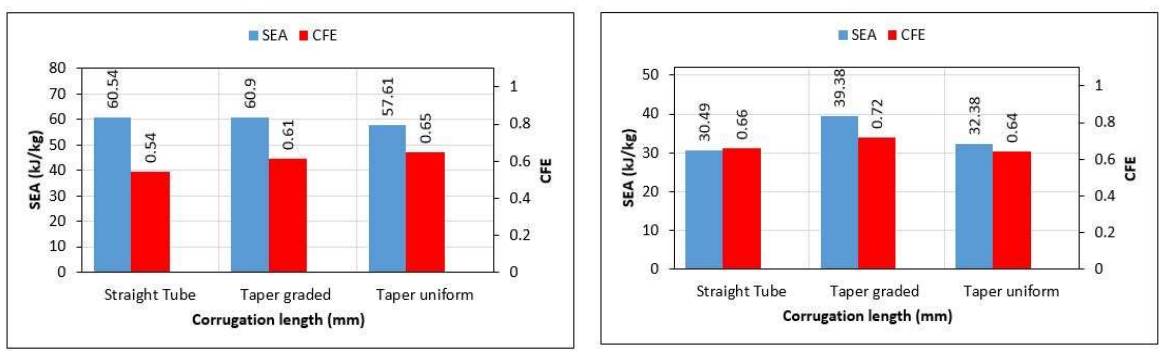

Fig 21. SEA and CFE of (a) empty straight and tapered tubes, (b) Foam filled straight and tapered tubes

The total energy absorption values of empty, foam filled straight and tapered tubes with uniform and graded thickness is shown in table 3. The maximum and minimum total energy absorption of empty tubes were found to be 9.38 kJ with tapered graded and 7.72 kJ in uniform tapered tube. Fig 21(a) shows the highest SEA of 60.9 kJ/kg was observed in tapered graded thickness and lowest in 57.61 kJ/kg. The maximum and minimum CFE values were found to be 0.61 in graded tube and 0.54 in uniform thickness tapered tube respectively. In foam filled case, the maximum and minimum total energy absorption were found to be 12.72 kJ/kg in graded taper and 9.91 kJ/kg uniform taper tubes respectively. The SEA of both foam filled tapered tube exceeds the straight tube as shown in Fig 21(b). The highest and lowest SEA values were 39.38 kJ/kg in graded and 30.49 kJ/kg in straight circular tube. The maximum CFE value, 0.72 was found in graded and minimum value, 0.64 was found in uniform taper tube respectively. Overall, in both empty and foam filled cases the tapered graded thickness tube having higher energy absorption capacity, SEA followed by straight circular then followed by tapered uniform tube.

## *5.5 Influence of amplitude and wavelength of corrugated tubes on the crashworthy parameters*

Fig. 22 shows the trend of $F_{max}$ with different amplitude and wavelengths, where the $F_{max}$ increases with the wavelength and the maximal $F_{max}$ reaches 136.56 kN (Empty) and 152.31 kN (foam filled), which is lower than that of straight circular tube 140 kN (Empty) and 157.42 kN (foam filled). Overall the $F_{max}$ increases with increase in wavelength and decrease in amplitude. By introducing corrugations to a straight tube, the crashing peak force $F_{max}$ can be significantly reduced.

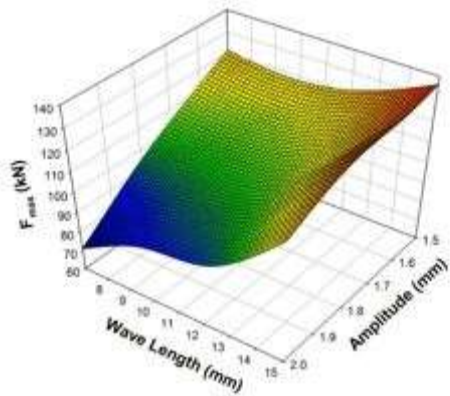 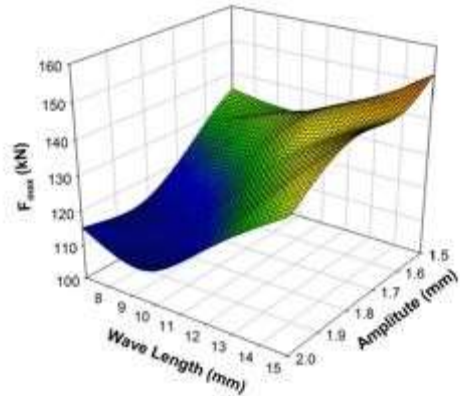

(a) Empty tube      (b) Foam filled tube

Figure 22. Influence of corrugation wavelength and amplitude on the $F_{max}$

Fig. 23 shows the trend of total energy absorption (EA) with different amplitude and wavelengths. The EA of empty and foam filled straight tube are 7.72 kJ and 11.04 kJ respectively. The highest EA of empty and foam filled corrugated tube are 6.99 kJ and 9.85 kJ. It shows that straight tube is more advantageous than corrugated tube in terms of EA. This implies that to keep a stable concertina mode of deformation, reduce the $F_{max}$ value corrugated tube must sacrifice some of its total energy absorption capacity. The EA increases with increase in corrugation wavelength. When wavelength increases, corrugated tube tends to be straight, which enhances the load carrying capacity of structure, which is the reason for a higher EA at higher wavelength. The EA decreases with increase in amplitude of corrugated tube.

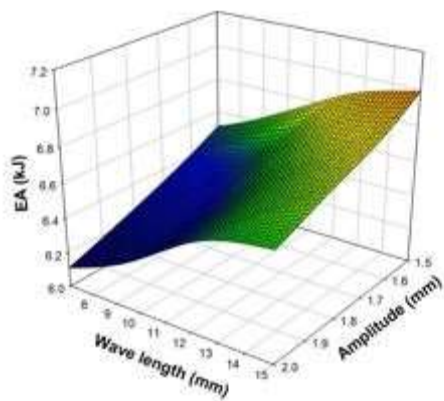 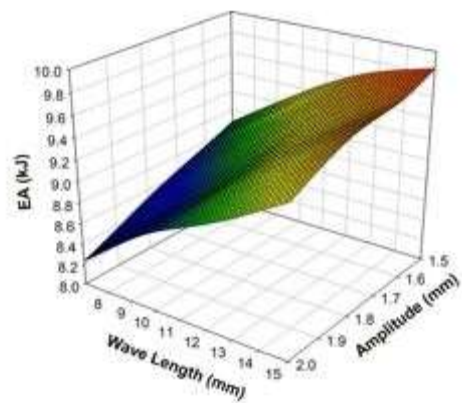

(a) Empty tube      (b) Foam filled tube

Figure 23. Influence of corrugation wavelength and amplitude on the EA

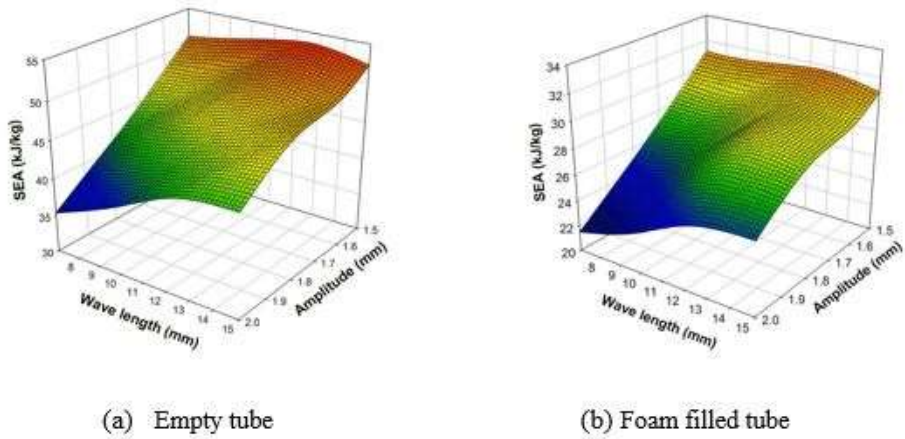

Figure 24. Influence of corrugation wavelength and amplitude on the SEA

Fig. 24 shows Influence of corrugation wavelength and amplitude on the SEA. The structural mass of empty and foam corrugated tube decrease with the wavelength. SEA of corrugated tube increases with increase in wavelength and decrease in amplitude.

## 6. Conclusion

In this paper, a comparative study on the dynamic crushing behaviour of straight tubes, tapered tubes and corrugated tubes were performed. A wide range of crashworthiness parameters, such as mode of deformation, initial peak force, mean force, total energy absorption, specific energy absorption, and crash force efficiency have been analyzed. The study also investigated the influence of amplitude and corrugation wave length of corrugated tubes on crashworthiness characteristics. From this study, it can be concluded that:

- The impact crushing behaviour of corrugated tubes is affected by its corrugation wavelength and amplitude.

- Corrugation on tube surface reduces the initial peak force and fluctuation in the force displacement diagram.
- Corrugation enhances the collapse mode of aluminum tubes. Corrugation tubes with wavelength below 7mm guarantee the tube to collapse in the Concertina mode, which is favourable for crashworthy application.
- The initial peak force and specific energy absorption of corrugation tubes increases with increase in corrugation wavelength and decrease in amplitude.
- Highest SEA was found in empty corrugated tube of corrugation wavelength 7 mm and amplitude 2mm.
- CFE values of foam-filled tubes are higher than that of empty tubes in types of tube configurations.
- Tapered tubes with graded thickness is superior to uniform thickness in terms of crashworthy parameters.